\documentclass[11pt]{article}

\usepackage{amsmath}
\usepackage{amsthm}
\usepackage{amssymb}
\usepackage{graphicx}
\usepackage{epsfig}

\newtheorem{Theorem}{Theorem}[section]

\newtheorem{Lemma}[Theorem]{Lemma}

\theoremstyle{definition}
\newtheorem{Definition}{Definition}[section]

\def\Ack{\medskip\noindent {\bf Acknowledgements:}\ \ignorespaces}

\def\signra{\bigskip \begin{center}
{\sc Ricardo Alonso\par\vspace{3mm}
Department of Mathematics, University of Texas at Austin \\
Austin, TX 78712, U.S.A.
\par\vspace{3mm}
e-mail:} \tt{ralonso@math.utexas.edu}  \end{center}}

\begin{document}

%\pagenumbering{arabic}

\title{Existence of global solutions to the Cauchy problem for the inelastic Boltzmann equation with near-vacuum data}

\author{Ricardo J. Alonso\footnote{Department of Mathematics, University of Texas at Austin.  Partial support from NSF under grant DMS-0507038.}}

\maketitle

%%%%%%%%%%%%%%%%%%%%%%%%%%%%%%%%%%%%%%%%%%%%%%%%%%%%%%%%%%%%%%%%%%%%%%%%%%%%%%%%%%%%%
%%%%%%%%%%%%%%%%%%%%%%%%%%%%%%%%%%%% Abstract %%%%%%%%%%%%%%%%%%%%%%%%%%%%%%%%%%%%%%%
%%%%%%%%%%%%%%%%%%%%%%%%%%%%%%%%%%%%%%%%%%%%%%%%%%%%%%%%%%%%%%%%%%%%%%%%%%%%%%%%%%%%%

\begin{abstract}
The Cauchy problem for the inelastic Boltzmann equation is studied for small data.  Existence and uniqueness of mild and weak solutions is obtained for sufficiently small data that lies in the space of functions bounded by Maxwellians.  The technique used to derive the result is the well known iteration process of Kaniel $\&$ Shinbrot.
\end{abstract}

%%%%%%%%%%%%%%%%%%%%%%%%%%%%%%%%%%%%%%%%%%%%%%%%%%%%%%%%%%%%%%%%%%%%%%%%%%%%%%%%%
%%%%%%%%%%%%%%%%%%%%%%%%%%%%%%%%%%% Introduction %%%%%%%%%%%%%%%%%%%%%%%%%%%%%%%%
%%%%%%%%%%%%%%%%%%%%%%%%%%%%%%%%%%%%%%%%%%%%%%%%%%%%%%%%%%%%%%%%%%%%%%%%%%%%%%%%%

\section{Introduction}
The theory developed by DiPerna $\&$ Lions in the 90's \cite{Di} on what is called \textsl{Renormalized solutions} has been a great success in finding existence theorems for the Boltzmann equation (BE):  The Cauchy problem in \cite{Di} and the Boundary value problem \cite{Ha}. The theory is strong and flexible and can be adapted to find solutions for different problems, for instance: The Vlasov-Poisson-Boltzmann system (VPB) \cite{Mi}, the treatment of the BE with infinite energy \cite{MiP}, the relativistic Boltzmann equation \cite{D} and others.  Indeed, a great deal of applications of this theory has been written in the last 18 years on BE related problems.\\\\
The theory is based in the bounded entropy which is a feature of the elastic BE solutions.  Unfortunately, it is not known how to obtain an a priori estimate that confirms such feature for the inelastic BE solutions, even in the case where a cut-off is imposed to the collision kernel.  This simple fact creates a big upset for the theory in the inelastic case.  It is important to say that in the one dimensional case, Benedetto $\&$ Pulvirenti overcame this problem in \cite{Be} provided that the initial datum is essentially bounded and has compact support in the velocity space.  The proof uses an iterative process introduced by J. M. Bony that relies strongly on the dimension.  The technique allows to find a uniform control on the entropy, and hence, to prove existence and uniqueness provided the initial datum has the afore mentioned properties.\\\\
It is clear that more understanding in the collision operator is needed to solve the inelastic BE in its simplest form: The Cauchy problem.  More complex problems, like the initial/boundary value problem or the VPB system, are still out of hand.  This paper returns to the late 70's and presents an application for the inelastic Boltzmann problem of the technique introduced by Kaniel $\&$ Shinbrot in that time.  Known as Kaniel $\&$ Shinbrot iterates \cite{Ka}, this technique was created by these authors to find existence and uniqueness of solutions for the BE in sufficiently small time.  The argument is beautiful and simple, however, the lack of a priori estimates makes difficult to use it for large time existence, particularly for hard potentials (see \cite{An} for an interesting result).  We refer the work of Ukai $\&$ Asano \cite{UA} for a successful extension from small to large time existence using Semigroup theory in the case of soft potentials.\\\\
In the mid 80's, Illner \& Shinbrot~\cite{Il} realized that it is possible to make a trade-off between the ``size" of the initial datum and the size of the time interval where the existence of solutions can be proved.  They modified Kaniel $\&$ Shinbrot argument to obtain global in time solutions in the elastic case.  Indeed, Illner \& Shinbrot~\cite{Il} observed that if the initial datum was close to vacuum and had sufficient decay at infinity, the argument could be carried out after dominating the solution globally in time with appropriate estimates.  See also Palczewski $\&$ Toscani~\cite{PT} for an additional interesting application of Kaniel $\&$ Shinbrot iteration on the existence of the BE when the initial data is close to a local Maxwellian.\\\\ 
Since the work of Illner \& Shinbrot~\cite{Il}, an extensive study has been made in the cases when the initial datum is small or a perturbation of the vacuum.  Regularity, asymptotic trend, stability and quantitative properties of solutions can be found in \cite{Bell}, \cite{Bou}, \cite{H}, \cite{Ha0}, and \cite{To}.  Recently, Glassey \cite{Gl} was able to find such global in time estimates for the relativistic case using the space
\begin{displaymath}
\left\{f\in C^{0}:\left\|(1+\left|x\times\xi\right|^{2})^{(1+\delta)/2}f\right\|_{\infty}<\infty\right\}.
\end{displaymath}
The weight $(1+\left|x\times\xi\right|^{2})^{(1+\delta)/2}$, more appropriate for the relativistic case, replaces the Maxwellian weight used originally by Kaniel \& Shinbrot (see S.-Y. Ha \cite{H} for different polynomial weights and $L^{1}$-stability).  The parameter $0<\delta<1$ is related to the decay hypothesis imposed in the scattering cross section.  It is important to observe that the advection coefficient in the relativistic case is bounded, this fact makes this case more handleable.  However, the higher complexity of the collision laws introduces great difficulty in the calculations of the estimates.\\\\
Finally, it is valuable to mention that for the inelastic homogeneous BE considerable work has been done, see \cite{BC}, \cite{BCT}, \cite{BCG}, \cite{BGP}, \cite{CIS}, \cite{EP}, \cite{GPV} and \cite{MMR} for extensive studies.  In the Maxwellian molecules case (the relative velocity in the collision frequency is approximated by the thermal speed) the reader may go to the work of Bobylev, Carrillo \& Gamba~\cite{BCG} and investigate different aspects of this problem, for example, existence and uniqueness of solutions, self-similar solutions and moment equations. In the hard sphere case, one of the first complete studies of inelastic interactions was done by Gamba, Panferov \& Villani in \cite{GPV}.  In this work the authors investigate the BE for diffusively excited granular media, i.e. they study the equation
\[
\partial_{t}f-\mu\Delta_{\xi}f=Q(f,f).
\]
Under precise conditions they find existence of a solution that becomes rapidly decaying and smooth for arbitrary small time.  It is also proved that such solution has overpopulated high-velocity tails with respect to the Maxwellian distribution.  In fact, they prove that the solution is controlled from below by
\[
K\exp(-a|\xi|^{3/2}).
\]
For an extensive study of general hard spheres kernels in the inelastic homogeneous case, we refer the work of Mischler, Mouhot \& Ricard \cite{MMR}.  In this paper the authors study stability, existence and uniqueness for the inelastic homogeneous BE with general hard sphere collision rate.  In addition, they investigate the long-time behavior of solutions, in particular, they show that solutions of the inelastic homogeneous BE problem collapse in the weak$*$ measure sense to the Dirac mass after (and not before) infinite time.  Therefore, a total loss of energy in the system occurs as $t\rightarrow\infty$.  This collapse phenomenon was first observed in the late 90's by Benedetto \& Caglioti~\cite{BEC} in a one-dimensional system of $n$-inelastic particles with constant restitution coefficient.  It is interesting that this collapse does not occur in the Maxwellian molecules case when variable restitution coefficient with appropriate behavior for small relative velocities is used, see \cite{BCG} for a detailed discussion.  In \cite{BGP} Bobylev, Gamba \& Panferov develop a complete study of the high-energy asymptotic (energy tails) for different inelastic regimes with forcing or heating terms.  They complete this study by means of the Povzner-type inequalities technique.\\\\
In general, all these previous studies of the space homogeneous problem are achieved using the weak formulation of the BE as the starting point, such approach is different from the one followed in this work for the space inhomogeneous case which uses the ``mild" formulation of the BE.  In addition, the difficulty introduced by the advection term $\xi\cdot\nabla f$ in the full BE used in the space inhomogeneous case will require a near vacuum assumption in our approach.  Such assumption is not necessary for the homogeneous case that only requires an initial datum with mass and first two moments finite to find existence and uniqueness of solutions.\\\\
This paper is organized as follows: Section 2 is devoted to present the problem and the notation.  Section 3 presents the Kaniel $\&$ Shinbrot iterates with small adaptations to the present case.  Section 4 shows different estimates essential to find global solutions, and, section 5 presents the existence theorems and some conclusions.

%%%%%%%%%%%%%%%%%%%%%%%%%%%%%%%%%%%%%%%%%%%%%%%%%%%%%%%%%%%%%%%%%%%%%%%%%%%%%%%%%
%%%%%%%%%%%%%%%%%%%%%%%%%%%%inelastic boltzmann equation%%%%%%%%%%%%%%%%%%%%%%%%%
%%%%%%%%%%%%%%%%%%%%%%%%%%%%%%%%%%%%%%%%%%%%%%%%%%%%%%%%%%%%%%%%%%%%%%%%%%%%%%%%%

\section{Inelastic Boltzmann Equation}
\subsection{The Cauchy Problem}
Let us assume that we have a large space filled with particles that are considered as mass points.  Assume that these particles are interacting with a specific law and that the particles are not influenced by external forces.  A good model to represent such dynamical system is given by the equation
\begin{equation}\label{E1IBE}
\frac{\partial{f}}{\partial{t}}+\xi\cdot\nabla{f}=Q(f,f)\;\;\mbox{in}\;\;(0,+\infty)\times\mathbb{R}^{n}\times\mathbb{R}^{n}
\end{equation}
The function $f(t,\xi,x)$, where $(t,x,\xi)\in(0,\infty)\times\mathbb{R}^{n}\times\mathbb{R}^{n}$, represents for any fix time $t$ and velocity $\xi$ the distribution of the density of particles throughout space $x$, for the fix time $t$ and the fix velocity $\xi$.  Thus, the physical meaning implies that $f\geq0$.  Equation (\ref{E1IBE}) is known as the Boltzmann equation because it was derived by the first time by L. Boltzmann in 1872 in his studies of dilute gases.  Today, the BE is used to model not only dilute gases but statistical transport associated to dynamical systems that behave ``like" a dilute gas.  Some examples are large scale interactions of Galaxies, granular gases and chemical reaction gases.\\\\
The Cauchy BE problem consist in finding a nonnegative function $f$ such that it solves the equation (\ref{E1IBE}) and the initial condition
\begin{equation}\label{IC}
f(0,\xi,x)=f_{0}(\xi,x)\;\;\mbox{on}\;\;\{0\}\times\mathbb{R}^{n}\times\mathbb{R}^{n}
\end{equation}
for some nonnegative function $f_{0}$.  Note that the BE is a linear transport equation equated to a term $Q(f,f)$.  This term is called collision operator and its purpose is to model the interaction or intramolecular interaction between particles.\\\\
Some notation is needed before the collision operator is introduced.  Write $\xi,\xi_{*}$ for the velocities of two particles just before they collide.  The symbols $\xi',\xi'_{*}$ are used for the velocities of these particles just after the collision occurs.  The law that relates pre-collision and post-collision velocities is given by
\begin{equation}\label{E2IBE}
\xi'=\xi-\frac{1+e}{2}(u\cdot{n})n\;\;\mbox{and}\;\;\xi'_{*}=\xi_{*}+\frac{1+e}{2}(u\cdot{n})n.
\end{equation}
The variable $u$ is used for the relative velocity between the particles
\begin{displaymath}
u=\xi-\xi_{*},
\end{displaymath}
and the unit vector $n\in S^{n-1}$ determines the impact direction, i.e. the unit vector that points from the $\xi$-particle center to the $\xi_{*}$-particle center at the moment that the particles collide.  The parameter $e$ is called the restitution coefficient.  The purpose of $e$ is to describe the inelastic effect in the collision between particles.  A good physical approximation is to take $e$ as a function of the impact velocity, i.e. a function of $|u\cdot n|$.  Contrary to the elastic case ($e$=1), in the inelastic case ($e<1$) the vector $n$ does not bisects the angle between the pre-collision and post-collisional relative velocities. In this work it is assumed that the restitution coefficient is \textsl{only} a function of the impact velocity $e=e(|u\cdot n|)$.  The properties of the map $e:z\rightarrow e(z)$ will be stated carefully later on.\\\\
In addition to equations (\ref{E2IBE}), there are other ways to parameterize the interaction between two particles, see for example \cite{BGP}, \cite{MMR} for complete studies using the center of mass-relative velocity parameterization.  However, the impact direction parameterization given by (\ref{E2IBE}) is appropriate for the present work since it shows the dependence of the impact velocity explicitly.\\\\
The notation with acute marks extends naturally to the pre-collision perspective using the symbols $'\!\xi,'\!\xi_{*}$ to denote the pre-collision velocities of a pair of particles.  For example, from the pre-collision perspective, equations (\ref{E2IBE}) read
\begin{equation}\label{E2.5IBE}
\xi={'\!\xi}-\frac{1+{'\!e}}{2}({'\!u}\cdot{n})n,\;\;\;\xi_{*}={'\!\xi_{*}}+\frac{1+{'\!e}}{2}({'\!u}\cdot{n})n
\end{equation}
where ${'\!e}=e(|{'\!u}\cdot n|)$.  Observe that in equations (\ref{E2.5IBE}) the symbols $\xi,\xi_{*}$ represents the post-collision velocities of the particles.\\\\
The Jacobian of the transformation (\ref{E2IBE}) will be needed subsequently for the manipulation of the collision operator. It is not hard to realize that
\begin{equation}\label{E2.7IBE}
J\left(|u\cdot n|\right)=\left|\frac{\partial\xi',\xi'_{*}}{\partial\xi,\xi_{*}}\right|=e\left(|u\cdot n|\right)+|u\cdot n|\;e_{z}\left(|u\cdot n|\right)=\theta_{z}\left(|u\cdot n|\right),
\end{equation}
where $\theta:z\rightarrow z\;e(z)$.  The symbols $e_{z}(z)$ and $\theta_{z}(z)$ have been chosen to denote the derivative of $e$ and $\theta$ respectively, avoiding the possible confusion with the post-variable symbols ``$e'(z)$" and ``$\theta'(z)$".  In this way, the Jacobian is fully determined by the restitution coefficient $e$ and hence, the Jacobian is a function only of the impact velocity $|u\cdot n|$.  Consequently, we denote $'\!J=J(|'\!u\cdot n|)$ and likewise for the symbols $e'$ and  $J'$.\\\\
After agreed in the previous notation let us describe the explicit form of the collision operator.  For this purpose fix $f$ and $g$ two nonnegative functions, then $Q(f,g)$ can be written as
\begin{equation}\label{E3IBE}
Q(f,g)=Q_{+}(f,g)-Q_{-}(f,g)
\end{equation}
where $Q_{+}(f,g)$ is known as the positive or gain part of the collision operator and is given by the integral formula
\begin{equation}\label{E4IBE}
Q_{+}(f,g)=\int_{\mathbb{R}^{n}}\int_{S^{n-1}}\frac{1}{'\!e\;'\!J}f('\!\xi)g('\!\xi_{*})|u\cdot n|dnd\xi_{*}.
\end{equation}
The term $Q_{-}(f,g)$ is referred as the negative or loss part of the collision operator and is given by the integral formula
\begin{equation}\label{E5IBE}
Q_{-}(f,g)=\int_{\mathbb{R}^{n}}\int_{S^{n-1}}f(\xi)g(\xi_{*})|u\cdot n|dnd\xi_{*}.
\end{equation}
An important point to observe is that $Q_{-}(f,g)$ can be written in the simpler form
\begin{displaymath}
Q_{-}(f,g)=f\cdot R(g),
\end{displaymath}
where
\begin{equation}\label{E6IBE}
R(g)=\int_{\mathbb{R}^{n}}\int_{S^{n-1}}g(\xi_{*})|u\cdot n|dn d\xi_{*}=C_{n}\int_{\mathbb{R}^{n}}g(\xi_{*})|u|d\xi_{*}=C_{n}\;g\ast|\xi|.
\end{equation}
Note that the constant $C_{n}$ can be easily calculated as,
\begin{equation}\label{E6.1IBE}
C_{n}=\int_{S^{n-1}}|\hat{u}\cdot n|dn=2\left|S^{n-2}\right|\int^{1}_{0}z(1-z^{2})^{(n-3)/2}dz=\frac{2}{n-1}\left|S^{n-2}\right|,
\end{equation}
in particular for the three dimensional case one has that $C_{3}=2\pi$.\\\\
$Remark:$ In the definition of the collision operator $Q(f,g)$, the hard sphere kernel $|u\cdot n|$ has been used.  This is common in the modeling of inelastic granular flows since in these, the interaction between particles is in fact a physical collision, i.e. the particles are modeled as if they were hard spheres colliding.
\subsection{Restitution coefficient}
The assumptions on the restitution coefficient $e$ are:
\begin{description}
\item [\it (A1)] $z\rightarrow e(z)$ is absolute continuous from $[0,+\infty)$ into $(0,1]$.
\item [\it (A2)] The mapping $z\rightarrow \theta(z)=ze(z)$ is strictly increasing.  Thus, $\theta(z)$ is invertible.
\item [\it (A3)] Define for any $\beta>0$ the ratio
\[
\psi_{\beta}(z)=\frac{\exp\left(-\frac{\beta}{2}\;z^{2}\right)_{z}}{\exp\left(-\frac{\beta}{2}\;\theta^{2}\right)_{z}}.
\]
Then, the following integrability condition on $\psi_{\beta}$ must hold
\begin{equation}\label{RC1}
\phi_{\beta}(x)=2\left|S^{n-2}\right|\int^{1}_{0}\psi_{\beta}(xz)\left(1-z^{2}\right)^{\frac{n-3}{2}}dz\in L^{\infty}([0,+\infty)).
\end{equation}
\end{description}
The elastic case is obtained when $e(z)\equiv1$.  Assumption \textsl{(A1)} is natural since physical restitution coefficients behave in a continuous fashion.  In addition, this assumption allows computing a.e. the Jacobian of the transformation (\ref{E2.5IBE}) using formula (\ref{E2.7IBE}).\\\\
Assumption \textsl{(A2)} implies, by equation (\ref{E2.7IBE}), that the Jacobian is positive.  Thus, the transformation (\ref{E2.5IBE}) is invertible.  Indeed, subtracting equations (\ref{E2.5IBE})
\begin{equation}\label{E7IBE}
u'\cdot n=-e(|u\cdot n|)\;(u\cdot n),
\end{equation}
hence, it is straightforward to show that
\begin{equation}\label{E8IBE}
u'\cdot n=-\mbox{sgn}(u\cdot n)\theta(|u\cdot n|)\;\;\mbox{and}\;\;u\cdot n=-\mbox{sgn}(u'\cdot n)\theta^{-1}(|u'\cdot n|).
\end{equation}
The explicit expressions of $\xi(\xi',\xi'_{*})$ and $\xi_{*}(\xi',\xi'_{*})$ are immediate from equations (\ref{E2IBE}) with equations (\ref{E8IBE}) at hand.  Also, observe that (\textsl{A2}) imposes an asymptotic pointwise trend on the restitution coefficient.  Indeed, note that  
\begin{displaymath}
e(z)=\frac{\theta(z)}{z}\geq\frac{\theta(1)}{z}=\frac{e(1)}{z}\;\;\mbox{for}\;z\geq1,
\end{displaymath}
thus, the decay of the restitution coefficient is controlled by below with the function $1/z$.\\\\
Condition (\textsl{A3}) is an integrability condition imposed indirectly on the Jacobian and it is required for technical purposes.  Although it may look strange, it is virtually satisfied in any practical case (for dimension $n\geq 2$).  It is certainly satisfied trivially in the elastic case and the following examples of commonly used restitution coefficients. 
\begin{description}
\item [\it (1)]$e(z)=e_{0}\in (0,1]$.  Constant restitution coefficient.
\item [\it (2)]$e(z)=\frac{1}{1+a\;z^{\gamma}}$, where $a>0$ and $\gamma\in(0,1]$.  Monotonic decreasing restitution coefficient.
\item [\it (3)]Continuous restitution coefficient that is elastic for small velocities and inelastic for large velocities.
\begin{displaymath}
e(z)=\left\{\begin{array}{cc} 1 & \mbox{for}\;z<z_{0} \\ e_{0}(z) & \mbox{for}\;z\geq z_{0}\end{array}\right. \ \ \mbox{for some} \ \ z_{0}>0,
\end{displaymath}
where $e_{0}(z)$ is usually a smooth decreasing function that may go to zero as $z\rightarrow\infty$.
\item [\it (4)]Viscoelastic hard spheres (Brilliantov $\&$ P\"oschel \cite{BP}).  In this case the dependence between the restitution coefficient and the impact velocity is given approximately by the equation\footnote{The restitution coefficient for viscoelastic hard spheres can be described as a power series of $z^{1/5}$ as well, i.e. $e(z)=\sum_{k}a_{k}\;z^{k/5}$, for details see~\cite{BP}.}
\begin{equation}\label{ve}
e(z)+a\;z^{1/5}e(z)^{3/5}=1,
\end{equation}
where the parameter $a\geq0$ (being $a=0$ the elastic case) is a constant depending on the material viscosity.  Let us verify that such restitution coefficient fulfill assumptions (\textsl{A2}) and (\textsl{A3}).  Differentiating equation (\ref{ve}) with respect to $z$ one obtains
\begin{displaymath}
ze_{z}(z)=-\frac{a/5\;z^{1/5}e(z)^{3/5}}{1+3a/5\;z^{1/5}e(z)^{-2/5}},
\end{displaymath}
therefore,
\begin{displaymath}
\theta_{z}(z)=\frac{1+2a/5\;z^{1/5}e(z)^{-2/5}}{1+3a/5\;z^{1/5}e(z)^{-2/5}}\;e(z)>0 \ \ \mbox{condition (\textsl{A2})}.
\end{displaymath}
In addition, for any $\beta>0$,
\begin{equation}\label{ve1}
\psi_{\beta}(z)=\frac{z}{\theta_{z}(z)}\exp\left(-\frac{\beta}{2}\;(z^{2}-\theta(z)^{2})\right).
\end{equation}
From equation (\ref{ve}) follows that $e(z)\rightarrow 0$ as $z\rightarrow\infty$, therefore
\begin{displaymath}
z^{2}-\theta(z)^{2}=\left(1-e(z)^{2}\right)z^{2}\sim z^{2}\ \ \mbox{for large}\;z,
\end{displaymath}
this suffices to prove that $\psi_{\beta}\in L^{\infty}\left([0,+\infty)\right)$, which implies assumption (\textsl{A3}) for dimension $n\geq 2$.
\end{description}

%%%%%%%%%%%%%%%%%%%%%%%%%%%%%%%%%%%%%%%%%%%%%%%%%%%%%%%%%%%%%%%%%%%%%%%%%%%%%%%%%%%%%%%%%%%%%%%%%%%%%%%%%%
%%%%%%%%%%%%%%%%%%%%%%%%%%%%%%%%%%%%%%%%%%%%%%%%%%%%%%%%%%%%%%%%%%%%%%%%%%%%%%%%%%%%%%%%%%%%%%%%%%%%%%%%%%

\subsection{Function spaces and properties of the collision operator}
The natural spaces to seek the solution of the Cauchy Problem are those functions in $L^{1}:=L^{1}(\mathbb{R}^{n}\times\mathbb{R}^{n})$ that are bounded by Maxwellians.
\begin{equation}\label{E1FS}
M^{\alpha,\beta}=\left\{f\in L^{1}:|f|\leq c\exp\left(-\alpha|x|^{2}\right)\exp\left(-\beta|\xi|^{2}\right)\;\;\mbox {for\;some}\;c>0\right\},
\end{equation}
where $\alpha$ and $\beta$ are positive parameters.  The notation $M^{0,\beta}$ is used when the function only has Maxwellian decay in the $\xi$ variable.\\
It is easy to check that the following one is a norm in these spaces
\begin{equation}\label{E2FS}
\left\|f\right\|_{\alpha,\beta}:=\inf_{c>0}\left\{|f|\leq c\exp\left(-\alpha|x|^{2}\right)\exp\left(-\beta|\xi|^{2}\right)\right\}.
\end{equation}
The notation adopted for the positive cone in $M^{\alpha,\beta}$ is
\begin{equation}\label{E3F2}
M^{\alpha,\beta}_{+}=\left\{f\in M^{\alpha,\beta}: f\geq0\right\}.
\end{equation}
The subscript $+$ is often used to denote the positive cone, thus, it will be used to denote the positive cone of any function space used in this work.\\\\
The Banach spaces $L^{p}(0,T;X)$ with $X$ a Banach space and $1\leq p\leq+\infty$ will be used to handle the time variable.  They are defined as the functions $f(t):[0,T]\rightarrow X$ that satisfy
\[
\left\|f\right\|_{L^{p}(0,T;X)}:=\left(\int^{T}_{0}\left\|f(t)\right\|^{p}_{X}\right)^{1/p}<+\infty\;\;\mbox{for}\;p<+\infty.
\]  
This quantity is, of course, a norm for these spaces.  The case $p=+\infty$ uses the same definition, with the integral changed for the essential supremum on $[0,T]$.  As an example, take $X=M^{\alpha,\beta}$ to obtain the spaces $L^{\infty}(0,T;M^{\alpha,\beta})$.  These are defined as
\begin{equation}
L^{\infty}(0,T;M^{\alpha,\beta})=\left\{f:[0,T]\rightarrow M^{\alpha,\beta}:\mbox{ess}\sup_{t\in[0,T]}\left\|f(t)\right\|_{\alpha,\beta}<\infty\right\}.
\end{equation}
Similarly, the standard spaces $C(0,T;X)$ and $W^{1,1}(0,T;X)$ are used.  Recall that $C(0,T;X)$ comprises all continuous functions $f:[0,T]\rightarrow X$ with the norm
\[
\left\|f\right\|_{C(0,T;X)}:=\max_{0\leq t\leq T}\left\|f(t)\right\|_{X}.
\]
Also, recall that $W^{1,1}(0,T;X)$ is defined as those functions $f\in L^{1}(0,T;X)$ such that $f_{t}$ exist in the weak sense and belongs to $L^{1}(0,T;X)$.  The norm used for this space is the obvious one
\[
\left\|f\right\|_{W^{1,1}(0,T;X)}=\left\|f\right\|_{L^{1}(0,T;X)}+\left\|f_{t}\right\|_{L^{1}(0,T;X)}.
\]
The collision operator defined by equations (\ref{E3IBE}),(\ref{E4IBE}) and (\ref{E5IBE}) enjoys some standard properties that are compiled here for the future use.\\
Fix $f,g\in M^{\alpha,\beta}_{+}$, then the gain and loss parts of the collision operator lie in $L^{1}_{+}$ and have the following properties:
\begin{description}
\item [\it ($P1_{a}$)] $\int_{\mathbb{R}^{n}}Q(f,f)(a+b\xi)d\xi=0$ for any $a,b\in\mathbb{R}$.  Conservation of mass and momentum.
\item [\it ($P1_{b}$)] $\int_{\mathbb{R}^{n}}Q(f,f)|\xi|^{2}d\xi\leq0$.  Loss of energy and cooling effect.
\item [\it ($P2$) ]$\left\|Q_{+}(f,g)\right\|_{L^{1}}=\left\|Q_{-}(f,g)\right\|_{L^{1}}$.
\item [\it ($P3$)]$\left\|Q_{-}(f,g)\right\|_{L^{1}}=\left\|Q_{-}(g,f)\right\|_{L^{1}}$.
\end{description}
The derivation of these properties is classical and follows after a change of variables $\{\xi,\xi_{*}\}\rightarrow\{'\!\xi,'\!\xi_{*}\}$ in the positive collision operator.  The reader may go to \cite{Ce} or \cite{Di} for the derivation in the elastic BE and to \cite{EP} for the derivation for the Enskog equation.

%%%%%%%%%%%%%%%%%%%%%%%%%%%%%%%%%%%%%%%%%%%%%%%%%%%%%%%%%%%%%%%%%%%%%%%%%%%%%%%%%
%%%%%%%%%%%%%%%%%%%%%%%%%%% linear problem %%%%%%%%%%%%%%%%%%%%%%%%%%%%%%%%%%%%%%
%%%%%%%%%%%%%%%%%%%%%%%%%%%%%%%%%%%%%%%%%%%%%%%%%%%%%%%%%%%%%%%%%%%%%%%%%%%%%%%%%

\section{Linear Problem}
\subsection{Mild solution}
As it will be explained later in more detail, the existence technique of Kaniel $\&$ Shinbrot for the Cauchy problem of the BE consists in finding the solution $f$ of the linear problem
\begin{equation}\label{E1LP}
\frac{\partial{f}}{\partial{t}}+\xi\cdot\nabla{f}+Q_{-}(f,g)=h\;\;\mathrm{with}\;\;f(0)=f_{0}
\end{equation}
with $g$ and $h$ carefully chosen, so that, $f$ approximates the actual BE solution.  Problem (\ref{E1LP}) is linear due to the bilinear character of the operator $Q$ and the fact that $g$ is fixed.\\\\
In order to solve problem (\ref{E1LP}) it is common to introduce the ``function along the trajectories",
\[
f^{\#}(t,x,\xi):=f(t,\psi_{t}(\xi,x)),
\]
where $\psi_{t}$ is the trajectory map of the transport operator $T:=\partial_{t}+\xi\cdot\nabla$
\[
\psi_{t}:(\xi,x)\longrightarrow(\xi,x+t\xi).
\]
Using this notation it is not hard to prove that if $f$ is smooth, equation (\ref{E1LP}) can be written equivalently as (recall that $Q_{-}(f,g)=f\cdot R(g)$ where $R(g)$ was given in equation (\ref{E6IBE})).
\begin{equation}\label{E2LP}
\frac{df^{\#}}{dt}(t)+f^{\#}R^{\#}(g)(t)=h^{\#}(t)\;\;\mathrm{with}\;\;f(0)=f_{0}.
\end{equation}
Equation (\ref{E2LP}) is a good base to define the concept of solution because it does not demand the differentiability in the $x$ variable for $f$, equation (\ref{E2LP}) does.  Moreover, if $f$ is smooth, equations (\ref{E1LP}) and (\ref{E2LP}) are equivalent in the sense that $f$ is a solution of the former if and only if is a solution of the later.  In other words, equation (\ref{E2LP}) is a generalization of equation (\ref{E1LP}).   
\begin{Definition}
Define \textsl{mild solution} in $[0,T]$ as a function $f\in W^{1,1}(0,T;L^{1})$, that solves a.e. equation (\ref{E2LP}) in $[0,T]\times\mathbb{R}^{n}\times\mathbb{R}^{n}$.
\end{Definition}
\noindent Observe that for any fix $t\geq 0$, the Jacobian of the trajectory map is 1,
\begin{displaymath}
\left|\frac{\partial\psi_{t}(\xi,x)}{\partial(\xi,x)}\right|=1.
\end{displaymath}
Thus, one has
\begin{Lemma}\label{L1LP}
Let $f\in L^{p}(0;T;L^{1})$ with $1\leq p\leq\infty$, then $\left\|f^{\#}(t)\right\|_{L^{1}}=\left\|f(t)\right\|_{L^{1}}$ for all $t\geq0$.
\end{Lemma}
%\begin{proof}
%It is clear that for any $t$, the Jacobian of the trajectory map is 1.
%\[
%\left|\frac{\partial\psi_{t}(\xi,x)}{\partial(\xi,x)}\right|=1.
%\]
%Thus,
%\begin{displaymath}
%\left\|f^{\#}(t)\right\|_{L^{1}}=\int_{\mathbb{R}^{n}}\int_{\mathbb{R}^{n}}|f(t,x+t\xi,\xi)|dxd\xi=\int_{\mathbb{R}^{n}}\int_{\mathbb{R}^{n}}|f(t,x,\xi)|dxd\xi=\left\|f(t)\right\|_{L^{1}}.
%\end{displaymath}
%\end{proof}
\noindent Note that the functions $f$ and $f^{\#}$ completely determine each other because the trajectory map $\psi_{t}$ is a bijection for all $t\geq 0$, therefore there is no ambiguity to express the subsequent results in terms of $f$ or $f^{\#}$.  The choice between the former and the later will be made so that the expressions are maintained as simple as possible.
\begin{Theorem}\label{L2LP}
Let $T>0$.  Let $f_{0}\in M^{\alpha,\beta}_{+}$, $g\in L^{\infty}(0,T;M^{0,\beta}_{+})$ and $h\geq0$ such that
\begin{equation}\label{E2.5LP}
\int^{t}_{0}h^{\#}(\tau)d\tau\in L^{\infty}(0,T;M^{\alpha,\beta}_{+}).
\end{equation}
Then (\ref{E2LP}) has a unique mild solution $f^{\#}\in L^{\infty}(0,T;M^{\alpha,\beta})$ given by the formula
\begin{equation}\label{E3LP}
f^{\#}(t)=f_{0}\;\exp\left(-\int^{t}_{0}R^{\#}(g)(\sigma)d\sigma\right)+\int^{t}_{0}h^{\#}(\tau)\exp\left(-\int^{t}_{\tau}R^{\#}(g)(\sigma)d\sigma\right)d\tau.
\end{equation}
Moreover, $\left\|f(t)\right\|_{L^{1}}$ is absolutely continuous, and
\begin{equation}\label{E4LP}
\frac{d}{dt}\left\|f(t)\right\|_{L^{1}}+\left\|Q_{-}(f,g)(t)\right\|_{L^{1}}=\left\|h(t)\right\|_{L^{1}},
\end{equation}
in particular $f\in C(0,T;L^{1}_{+})$.
\end{Theorem}
\begin{proof}
Since $g$ and $h$ are nonnegative, we may take nonnegative smooth sequences $\{g_{n}\}$ and $\{h_{n}\}$ such that
\[
g_{n}\nearrow g\;\;\mbox{and}\;\;h_{n}\nearrow h\;\;\mbox{p.w. as}\;n\rightarrow+\infty.
\]
Define the nonnegative sequence of smooth functions in the variable $t$
\begin{equation}\label{T1e1}
f^{\#}_{n}(t)=f_{0}\;\exp\left(-\int^{t}_{0}R^{\#}(g_{n})(\sigma)d\sigma\right)+\int^{t}_{0}h^{\#}_{n}(\tau)\exp\left(-\int^{t}_{\tau}R^{\#}(g_{n})(\sigma)d\sigma\right)d\tau.
\end{equation}
Clearly,
\[
f^{\#}_{n}\rightarrow f^{\#}\;\;\mbox{p.w. as}\;n\rightarrow+\infty,
\]
where $f^{\#}(t)$ is the nonnegative function given by equation (\ref{E3LP}).  Moreover,
\[
f_{0}\;\exp\left(-\int^{t}_{0}R^{\#}(g_{n})(\sigma)d\sigma\right)\leq f_{0}\in L^{1}(0,T;L^{1})
\]
and,
\[
\int^{t}_{0}h^{\#}_{n}(\tau)\exp\left(-\int^{t}_{\tau}R^{\#}(g_{n})(\sigma)d\sigma\right)d\tau\leq\int^{t}_{0}h^{\#}(\tau)d\tau\in L^{1}(0,T;L^{1}).
\]
Therefore, the convergence $f^{\#}_{n}\rightarrow f^{\#}$ is not only pointwise, is in fact in $L^{1}(0,T;L^{1})$.\\\\
In addition observe that since the sequence $\{f^{\#}_{n}(t)\}$ is smooth in time, by an elementary fact of ODE's
\begin{equation}\label{T1e2}
\frac{df^{\#}_{n}}{dt}(t)+f^{\#}_{n}R^{\#}(g_{n})(t)=h^{\#}_{n}(t)\;\;\mathrm{with}\;\;f_{n}(0)=f_{0}.
\end{equation}
Assumption (\ref{E2.5LP}) implies that $h^{\#}(t)\in L^{1}(0,T;L^{1})$.  Indeed, note that
\[
\int^{T}_{0}\left\|h^{\#}(\tau)\right\|_{L^{1}}d\tau=\left\|\int^{T}_{0}h^{\#}(\tau)d\tau\right\|_{L^{1}}\leq C_{\alpha,\beta}\left\|\int^{t}_{0}h^{\#}(\tau)d\tau\right\|_{L^{\infty}(0,T;M^{\alpha,\beta})},
\]
hence,
\[
h^{\#}_{n}(t)\rightarrow h^{\#}(t)\;\;\mbox{in}\;\;L^{1}(0,T;L^{1}).
\]
Similarly, after some easy computations one has the inequalities for $0\leq t\leq T$,
\[
R^{\#}(g_{n})(t)\leq R^{\#}(g)\leq C_{0}\left(\left\|g\right\|_{L^{\infty}(0,T;M^{0,\beta}_{+})}\right)(1+|\xi|)
\]
and
\begin{equation}\label{T1e3}
f^{\#}_{n}(t)\leq \left(\left\|f_{0}\right\|_{\alpha,\beta}+\;\left\|\int^{t}_{0}h^{\#}(\tau)d\tau\right\|_{L^{\infty}(0,T;M^{\alpha,\beta})}\right)\exp\left(-\alpha|x|^{2}\right)\exp\left(-\beta|\xi|^{2}\right)
\end{equation}
for some nonnegative constants $C_{0}$ and $C_{1}$ depending in the quantities as stated.  Thus,
\[
f^{\#}_{n}R^{\#}(g_{n})\rightarrow f^{\#}R^{\#}{g}\;\;\mbox{in}\;\;L^{1}(0,T;L^{1}).
\]
Equation (\ref{T1e2}) implies that
\[
\frac{df^{\#}_{n}}{dt}(t)\rightarrow\zeta\;\;\mbox{in}\;\;L^{1}(0,T;L^{1}),
\]
however, as it was said $f^{\#}_{n}\rightarrow f^{\#}$ in $L^{1}(0,T;L^{1})$, thus it is concluded that
\[
\zeta=\frac{df^{\#}}{dt}.
\]
Therefore, $f^{\#}$ and $f$ belong to $W^{1,1}(0,T;L^{1})$, i.e. $f$ is a mild solution of the linear problem (\ref{E2LP}).  Moreover, using (\ref{T1e3}) one concludes that $f^{\#}\in L^{\infty}(0,T;M^{\alpha,\beta})$.\\\\
For uniqueness, it suffices to prove that when $f_{0}$ and $h$ are zero, $f$ is also zero.  When $h=0$, problem (\ref{E2LP}) can be written as
\begin{displaymath}
\frac{d}{dt}\left(f^{\#}(t)\exp\left(\int^{t}_{0}R^{\#}(g)(\sigma)d\sigma\right)\right)=0
\end{displaymath}
Assuming also that $f_{0}=0$, follows directly from this equation that $f^{\#}=0$.\\\\
Finally, it is a well known fact (see Theorem~\ref{T2.5} in Appendix B), that for any real Banach space $X$ one has the inclusion $W^{1,1}(0,T;X)\subset C(0,T;X)$.  In particular, this inclusion is true for $X=L^{1}$.  Thus, integrating (\ref{E2LP}) in space and velocity and using the fact that
\[
\left\|\frac{df^{\#}}{dt}(t)\right\|_{L^{1}}=\frac{d}{dt}\left\|f^{\#}(t)\right\|_{L^{1}}=\frac{d}{dt}\left\|f(t)\right\|_{L^{1}},
\]
the equation (\ref{E4LP}) follows.  This concludes the proof.
\end{proof}
\subsection{Kaniel $\&$ Shinbrot Iteration}
Kaniel $\&$ Shinbrot method consists in building two sequences, $\{l_{n}\}$ and $\{u_{n}\}$, from the linear problem (\ref{E2LP}).  These sequences are built in such a way that one is monotone increasing while the other is monotone decreasing.  The key point is to prove that both squeeze down on a mild solution of the Boltzmann Equation.  The way to produce such sequences is the following:
Assume that $l_{0},\ldots,l_{n-1},u_{0}\ldots,u_{n-1}$ are known, then $l_{n}(t)$ and $u_{n}(t)$ are the mild solutions in $[0,T]$ of the linear problems
\begin{displaymath}
\frac{dl_{n}^{\#}}{dt}(t)+Q^{\#}_{-}(l_{n},u_{n-1})(t)=Q^{\#}_{+}(l_{n-1},l_{n-1})(t)
\end{displaymath}
\begin{equation}\label{E5LP}
\frac{du_{n}^{\#}}{dt}(t)+Q^{\#}_{-}(u_{n},l_{n-1})(t)=Q^{\#}_{+}(u_{n-1},u_{n-1})(t)
\end{equation}
with $l_{n}(0)=u_{n}(0)=f_{0}$.
The construction begins with a pair of functions $(l_{0},u_{0})$ satisfying what Kaniel $\&$ Shinbrot called \textsl{the beginning condition} in $[0,T]$, i.e. $u^{\#}_{0}\in L^{\infty}(0,T;M^{\alpha,\beta}_{+})$ and
\begin{equation}\label{E6LP}
0\leq l^{\#}_{0}(t)\leq l^{\#}_{1}(t)\leq u^{\#}_{1}(t)\leq u^{\#}_{0}(t)\;\;\mathrm{in}\;\; 0\leq t\leq T.
\end{equation}
\begin {Lemma}\label{L3LP}
Let $(l_{0},u_{0})$ satisfy the beginning condition in $[0,T]$.  Then, the sequences $\{l^{\#}_{n}\}$ and $\{u^{\#}_{n}\}$, defined in (\ref{E5LP}), exist for all $n$ and belong to $L^{\infty}(0,T;M^{\alpha,\beta}_{+})$.  Moreover these sequences satisfy
\begin{equation}\label{E7LP}
0\leq l^{\#}_{0}(t)\leq l^{\#}_{1}(t)\leq\ldots\leq l^{\#}_{n}(t)\leq\ldots\leq u^{\#}_{n}(t)\leq\ldots\leq u^{\#}_{1}(t)\leq u^{\#}_{0}(t)\;\;\mathrm{in}\;\ 0\leq t\leq T,
\end{equation}
consequently, $\{l^{\#}_{n}\}$ and $\{u^{\#}_{n}\}$ converge in $M^{\alpha,\beta}$ in $[0,T]$.
\end{Lemma}
\begin{proof}
The beginning condition implements the first step in the induction.  Next, assume $l^{\#}_{1},\cdots,l^{\#}_{k-1}$ and $u^{\#}_{1},\cdots,u^{\#}_{k-1}$ all exist, belong to $L^{\infty}(0,T;M^{\alpha,\beta}_{+})$ and satisfy for $t\in[0,T]$
\begin{equation}\label{E7.2LP}
0\leq l^{\#}_{0}(t)\leq\cdots\leq l^{\#}_{k-1}(t)\leq u^{\#}_{k-1}(t)\leq\cdots\leq u^{\#}_{0}(t).
\end{equation}
Since for any to functions $g$ and $f$
\begin{equation}\label{E7.3LP}
g\leq f\;\;\mathrm{if\;and\;only\;if}\;\;g^{\#}\leq f^{\#} 
\end{equation}
it is concluded that
\begin{equation}\label{E7.5LP}
0\leq Q^{\#}_{+}(l_{0},l_{0})\leq\cdots\leq Q^{\#}_{+}(l_{k-1},l_{k-1})\leq Q^{\#}_{+}(u_{k-1},u_{k-1})\leq\cdots\leq Q^{\#}_{+}(u_{0},u_{0}).
\end{equation}
Using the estimate of Lemma~\ref{L2ECO} in the next section,
\begin{displaymath}
\int^{t}_{0}Q^{\#}_{+}(u_{0},u_{0})(\tau)d\tau\in C(0,T;M^{\alpha,\beta}_{+}),
\end{displaymath}
therefore, using (\ref{E7.5LP})
\begin{displaymath}
\int^{t}_{0}Q^{\#}_{+}(l_{k-1},l_{k-1})(\tau)d\tau\;\;\mathrm{and}\;\;\int^{t}_{0}Q^{\#}_{+}(u_{k-1},u_{k-1})(\tau)d\tau
\end{displaymath}
lie in $C(0,T;M^{\alpha,\beta}_{+})$.  Moreover, it is clear that using (\ref{E7.3LP}) in the sequence of inequalities (\ref{E7.2LP})
\begin{displaymath}
0\leq l_{k-1}\leq u_{k-1}\leq \left\|u_{0}^{\#}\right\|_{L^{\infty}(0,T;M^{\alpha,\beta})}\exp\left(-\alpha|x-t\xi|^{2}-\beta|\xi|^{2}\right).
\end{displaymath}
Therefore $l_{k-1}$ and $u_{k-1}$ lie in $L^{\infty}(0,T;M^{0,\beta}_{+})$.  Applying Theorem~\ref{L2LP} to obtain that $l^{\#}_{k}$ and $u^{\#}_{k}$ exist and lie in $L^{\infty}(0,T;M^{\alpha,\beta}_{+})$.\\\\
Finally, observe that by (\ref{E7.2LP}) and (\ref{E7.3LP})
\begin{equation}\label{E7.7LP}
0\leq R^{\#}(l_{0})\leq\cdots\leq R^{\#}(l_{k-1})\leq R^{\#}(u_{k-1})\leq\cdots\leq R^{\#}(u_{0}).
\end{equation}
Using (\ref{E7.5LP}) and (\ref{E7.7LP}) one concludes directly from (\ref{E3LP}) that
\begin{displaymath}
0\leq l^{\#}_{k-1}\leq l^{\#}_{k}\leq u^{\#}_{k}\leq u^{\#}_{k-1},
\end{displaymath} 
this completes the proof.
\end{proof}
\noindent Since $\{l_{n}\}$ and $\{u_{n}\}$ are monotonic nonnegative sequences bounded by $u_{0}\in L^{\infty}(0,T;L^{1})$, they actually must converge in $L^{\infty}(0,T;L^{1})$ to some limits.  Moreover, one has 
\begin{Lemma}\label{L4LP}
Let $f_{0}\in M^{\alpha,\beta}_{+}$ and $(l_{0},u_{0})$ satisfy the beginning condition in $[0,T]$.  Denote the limits of $\{l_{n}\}$ and $\{u_{n}\}$ by $l(t)$ and $u(t)$.  Then $l(t)=u(t)$ for all $t\in[0,T]$.  Moreover this common limit is a mild solution of the Boltzmann equation
\begin{equation}\label{E8LP}
\frac{df^{\#}}{dt}(t)+Q^{\#}_{-}(f,f)(t)=Q^{\#}_{+}(f,f)(t)\;\;\mathrm{with}\;\;f(0)=f_{0}
\end{equation}
with $f^{\#}\in L^{\infty}(0,T;M^{\alpha,\beta}_{+})$.
\end{Lemma}
\begin{proof}
The fact that $l(t)=u(t)$ follows identically as in \cite{Ka}, Lemma 5.2.
It remains to show that the common limit is mild solution of (\ref{E8LP}).  Thus, let $f(t):=l(t)=u(t)$ and integrate equation (\ref{E5LP}) to obtain,
\begin{displaymath}
l_{n}^{\#}(t)+\int^{t}_{0}Q^{\#}_{-}(l_{n}(\tau),u_{n-1}(\tau))d\tau=f_{0}+\int^{t}_{0}Q^{\#}_{+}(l_{n-1}(\tau),l_{n-1}(\tau))d\tau.
\end{displaymath}
Send $n\rightarrow+\infty$ in this equation. Since the sequence $\{l_{n}^{\#}\}$ is increasing, it is possible to pass to the limit in the integrals to get an integral version of (\ref{E8LP}).  Thus, it is deduced that $f^{\#}\in W^{1,1}(0,T;L^{1})$, and by means of estimate (\ref{E3ECO}), $f^{\#}\in L^{\infty}(0,T;M^{\alpha,\beta}_{+})$.
\end{proof}
\begin{Lemma}\label{L5LP}
The mild solution $f^{\#}$ of the Boltzmann equation found in the Lemma~\ref{L4LP} is unique in the space $L^{\infty}(0,T;M^{\alpha,\beta}_{+})$.
\end{Lemma}
\begin{proof}
Assume that $f^{\#}_{1}$ and $f^{\#}_{2}$ are mild solutions of (\ref{E8LP}) that lie in $L^{\infty}(0,T;M^{\alpha,\beta})_{+}$.  Then for $i=1,2$,
\begin{equation}\label{E13LP}
f^{\#}_{i}(t)=\int^{t}_{0}Q^{\#}_{+}(f_{i},f_{i})(\tau)-Q^{\#}_{-}(f_{i},f_{i})(\tau)d\tau,\;\;\;\;f_{i}=f_{0}.
\end{equation}
Define $j(t):=f_{1}(t)-f_{2}(t)$.  After subtracting equations (\ref{E13LP}) and taking the $L^{1}$ norm in space and velocity,
\begin{align}
\nonumber\left\|j(t)\right\|_{L^{1}}&\leq\int^{t}_{0}\left(\left\|Q_{+}(f_{1},j)(\tau)\right\|_{L^{1}}+\left\|Q_{+}(j,f_{2})(\tau)\right\|_{L^{1}}+\left\|Q_{-}(f_{1},j)(\tau)\right\|_{L^{1}}+\left\|Q_{-}(j,f_{2})(\tau)\right\|_{L^{1}}\right)d\tau\\
&=2\int^{t}_{0}\left(\left\|Q_{-}(j,f_{1})(\tau)\right\|_{L^{1}}+\left\|Q_{-}(j,f_{2})(\tau)\right\|_{L^{1}}\right)d\tau.\label{E14LP}
\end{align} 
The inequality in the right hand side of (\ref{E14LP}) follows after applying properties (P2) and (P3).  However,
\begin{displaymath}
\left\|Q_{-}(j,f_{1})(\tau)\right\|_{L^{1}}+\left\|Q_{-}(j,f_{2})(\tau)\right\|_{L^{1}}\leq C_{\beta}\max_{i=1,2}\left\|f^{\#}_{i}\right\|_{\alpha,\beta}\left\|(1+|\xi|)j(\tau)\;\right\|_{L^{1}}.
\end{displaymath}
Using last inequality in (\ref{E14LP})
\begin{displaymath}
\left\|j(t)\right\|_{L^{1}}\leq C_{\beta}\max_{i=1,2}\left\|f^{\#}_{i}\right\|_{\alpha,\beta}\int^{t}_{0}\left\|(1+|\xi|)j(\tau)\;\right\|_{L^{1}}d\tau.
\end{displaymath}
Using Gronwall's Lemma one concludes that $j(t)=0$ in $[0,T]$.  We refer to \cite{Ka} for additional remarks about uniqueness.
\end{proof}

%%%%%%%%%%%%%%%%%%%%%%%%%%%%%%%%%%%%%%%%%%%%%%%%%%%%%%%%%%%%%%%%%%%%%%%%%%%%%%%%%%%
%%%%%%%%%%%%%%%%%%%%%%%%%%Estimates collision operator%%%%%%%%%%%%%%%%%%%%%%%%%%%%%
%%%%%%%%%%%%%%%%%%%%%%%%%%%%%%%%%%%%%%%%%%%%%%%%%%%%%%%%%%%%%%%%%%%%%%%%%%%%%%%%%%%

\section{Estimates on the Collision Operator}
\begin{Lemma}\label{L1ECO}
Let $'\!\xi$ and $'\!\xi_{*}$ defined by the pre-collision formulas (\ref{E2.5IBE}) then
\begin{equation}\label{E1ECO}
\int^{t}_{0}\exp\left(-\alpha\left|x+\tau(\xi-'\!\xi)\right|^{2}\right)\exp\left(-\alpha\left|x+\tau(\xi-'\!\xi_{*})\right|^{2}\right)d\tau\leq \frac{\sqrt{\pi}}{\alpha^{1/2}|u|}\exp\left(-\alpha\left|x\right|^{2}\right).
\end{equation} 
\end{Lemma}
\begin{proof}
Define the vector $b$ by
\begin{displaymath}
b=\xi-'\!\xi=-\frac{1+'\!e}{2}('\!u\cdot n)n
\end{displaymath}
therefore,
\begin{displaymath}
-b=\xi_{*}-'\!\xi_{*}=\frac{1+'\!e}{2}('\!u\cdot n)n.
\end{displaymath}
Thus, the following equality is obtained
\begin{multline}\label{E1.5ECO}
\left|x+\tau(\xi-'\!\xi)\right|^{2}+\left|x+\tau(\xi-'\!\xi_{*})\right|^{2}=\left|x+\tau b\right|^{2}+\left|x-\tau b+\tau u \right|^{2}=\\
\left|x\right|^{2}+\left|x+\tau u\right|^{2}+2\tau^{2}\left(\left|b\right|^{2}-b\cdot u\right).
\end{multline}
Now, since $u\cdot n=-'\!e('\!u\cdot n)$ one obtains,
\begin{equation}\label{E1.7ECO}
\left|b\right|^{2}-b\cdot u=\frac{1-'\!e^{2}}{4}\left|'\!u\cdot n\right|^{2}\geq 0.
\end{equation}
As a result of (\ref{E1.5ECO}) and (\ref{E1.7ECO}),
\begin{multline}\label{E2ECO}
\int^{t}_{0}\exp\left(-\alpha\left|x+\tau(\xi-'\!\xi)\right|^{2}\right)\exp\left(-\alpha\left|x+\tau(\xi-'\!\xi_{*})\right|^{2}\right)d\tau\leq\\\exp\left(-\alpha\left|x\right|^{2}\right)\int^{t}_{0}\exp\left(-\alpha\left|x+\tau u\right|^{2}\right)d\tau.
\end{multline}
Observe that,
\begin{displaymath}
\left|x+\tau u\right|^{2}=\left|x\right|^{2}-\left|x\cdot\hat{u}\right|^{2}+\left|u\right|^{2}\left(\frac{x\cdot\hat{u}}{|u|}+\tau\right)^{2}
\end{displaymath}
but $\left|x\right|^{2}-\left|x\cdot\hat{u}\right|^{2}\geq0$, then from the previous inequality 
\begin{align}
\int^{t}_{0}\exp\left(-\alpha\left|x+\tau u\right|^{2}\right)d\tau&\leq\int^{+\infty}_{-\infty}\exp\left(-\alpha\left|x+\tau u\right|^{2}\right)d\tau\nonumber\\
&\leq\int^{+\infty}_{-\infty}\exp\left(-\alpha\left|u\right|^{2}\left(\frac{x\cdot\hat{u}}{|u|}+\tau\right)^{2}\right)d\tau\nonumber\\
&=\int^{\infty}_{-\infty}\exp\left(-\alpha|u|^{2}\tau^{2}\right)d\tau=\frac{\sqrt{\pi}}{\alpha^{1/2}|u|}.\label{E2.9ECO}
\end{align}
Using inequality (\ref{E2.9ECO}) in (\ref{E2ECO}) one obtains (\ref{E1ECO}).
\end{proof}
\begin{Lemma}\label{L2ECO}
For any $0\leq t\leq T$ and $f^{\#}\in L^{\infty}(0,T;M^{\alpha,\beta})$ the following inequality holds
\begin{equation}\label{E3ECO}
\int^{t}_{0}\left|Q^{\#}_{+}(f,f)(\tau)\right|d\tau\leq k_{\alpha,\beta}\exp\left(-\alpha|x|^{2}-\beta|\xi|^{2}\right)\left\|f^{\#}\right\|^{2}_{L^{\infty}(0,T;M^{\alpha,\beta})},
\end{equation}
where
\[k_{\alpha,\beta}=C_{n}\frac{\left\|\phi_{\beta}\right\|_{L^{\infty}}}{\alpha^{1/2}\beta^{n/2}}.
\]
The constant $C_{n}$ only depends on the dimension.  In other words,
\begin{equation}\label{E3.1ECO}
\int^{t}_{0}\left|Q^{\#}_{+}(f,f)(\tau)\right|d\tau\in L^{\infty}(0,T;M^{\alpha,\beta}).
\end{equation}
\end{Lemma}
\begin{proof}
Note that
\begin{multline}
Q^{\#}_{+}(f,f)(\tau,x,\xi)=Q_{+}(f,f)(\tau,x+\tau\xi,\xi)=\\
\int_{\mathbb{R}^{n}}\int_{S^{n-1}}\frac{1}{'\!e\;'\!J}f^{\#}(\tau,x+\tau(\xi-'\!\xi),'\!\xi)f^{\#}(\tau,x+\tau(\xi-'\!\xi_{*}),'\!\xi_{*})|u\cdot n|dnd\xi_{*}
\end{multline}
hence,
\begin{multline}
\left|Q^{\#}_{+}(f,f)(\tau,x,\xi)\right|\leq
\left\|f^{\#}\right\|^{2}_{L^{\infty}(0,T;M^{\alpha,\beta})}\int_{\mathbb{R}^{n}}\int_{S^{n-1}}\frac{1}{'\!e\;'\!J}\exp\left(-\alpha|x+\tau(\xi-'\!\xi)|^{2}-\beta|'\!\xi|^{2}\right)\\\exp\left(-\alpha|x+\tau(\xi-'\!\xi_{*})|^{2}-\beta|'\!\xi_{*}|^{2}\right)|u\cdot n|dnd\xi_{*}.\label{E4ECO}
\end{multline}
Using the fact that
\begin{displaymath}
\exp\left(-\beta|'\!\xi|^{2}-\beta|'\!\xi_{*}|^{2}\right)=\exp\left(-\beta|\xi|^{2}-\beta|\xi_{*}|^{2}-\beta\frac{1-'\!e^{2}}{2}|'\!u\cdot n|^{2}\right),
\end{displaymath}
one gets from (\ref{E4ECO})
\begin{multline}
\left|Q^{\#}_{+}(f,f)(\tau,x,\xi)\right|\leq
\exp\left(-\beta|\xi|^{2}\right)\left\|f^{\#}\right\|^{2}_{L^{\infty}(0,T;M^{\alpha,\beta})}\int_{\mathbb{R}^{n}}\exp\left(-\beta|\xi_{*}|^{2}\right)\\\int_{S^{n-1}}\frac{1}{'\!e\;'\!J}\exp\left(-\beta\frac{1-'\!e^{2}}{2}|'\!u\cdot n|^{2}\right)\exp\left(-\alpha|x+\tau(\xi-'\!\xi)|^{2}\right)\\\exp\left(-\alpha|x+\tau(\xi-'\!\xi_{*})|^{2}\right)|u\cdot n|dnd\xi_{*}.\label{E5ECO}
\end{multline}
Integrating (\ref{E5ECO}) by $\tau$ and using Lemma~\ref{L1ECO} it follows that
\begin{multline}
\int^{t}_{0}\left|Q^{\#}_{+}(f,f)(\tau,x,\xi)\right|d\tau\leq\frac{\sqrt{\pi}}{\alpha^{1/2}}\exp\left(-\alpha|x|^{2}-\beta|\xi|^{2}\right)\left\|f^{\#}\right\|^{2}_{L^{\infty}(0,T;M^{\alpha,\beta})}\\
\int_{\mathbb{R}^{n}}\exp\left(-\beta|\xi_{*}|^{2}\right)\int_{S^{n-1}}\frac{1}{'\!e\;'\!J}\exp\left(-\beta\frac{1-{'\!e^{2}}}{2}|'\!u\cdot n|^{2}\right)|\hat{u}\cdot n|dnd\xi_{*}.\label{E6ECO}
\end{multline}
Observe that due to condition (\ref{RC1})
\begin{displaymath}
\int_{S^{n-1}}\frac{1}{'\!e\;'\!J}\exp\left(-\beta\frac{1-{'\!e^{2}}}{2}|'\!u\cdot n|^{2}\right)dn=\phi_{\beta}(|'\!u|)\leq \left\|\phi_{\beta}\right\|_{L^{\infty}}.
\end{displaymath}
As a result, it follows from (\ref{E6ECO}) that
\begin{displaymath}
\int^{t}_{0}\left|Q^{\#}_{+}(f,f)(\tau,x,\xi)\right|d\tau\leq C_{n}\frac{\left\|\phi_{\beta}\right\|_{L^{\infty}}}{\alpha^{1/2}\beta^{n/2}}\exp\left(-\alpha|x|^{2}-\beta|\xi|^{2}\right)\left\|f^{\#}\right\|^{2}_{L^{\infty}(0,T;M^{\alpha,\beta})},
\end{displaymath}
this proves the Lemma.
\end{proof}

%%%%%%%%%%%%%%%%%%%%%%%%%%%%%%%%%%%%%%%%%%%%%%%%%%%%%%%%%%%%%%%%%%%%%%%%%%%%%%%%%%%
%%%%%%%%%%%%%%%%%%%%%%%%%%Existence theorem%%%%%%%%%%%%%%%%%%%%%%%%%%%%%%%%%%%%%%%%
%%%%%%%%%%%%%%%%%%%%%%%%%%%%%%%%%%%%%%%%%%%%%%%%%%%%%%%%%%%%%%%%%%%%%%%%%%%%%%%%%%%

\section{Existence of Global Solution}
\begin{Theorem}\label{ET1}
Let $T>0$ and assume $f_{0}\in M^{\alpha,\beta}$ with 
\[
\left\|f_{0}\right\|_{\alpha,\beta}\leq\frac{1}{4k_{\alpha,\beta}}
\]
where $k_{\alpha,\beta}$ is given in estimate (\ref{E3ECO}).  Then, the problem (\ref{E8LP}) has a unique mild solution with $f^{\#}\in L^{\infty}(0,T;M^{\alpha,\beta}_{+})\cap C(0,T;L^{1}_{+})$.
\end{Theorem}
\begin{proof}
It suffices to prove that the beginning condition is satisfied globally.  Let
\begin{displaymath}
l^{\#}_{0}=0\;\;\;\;\mathrm{and}\;\;\;\;u^{\#}_{0}=C\exp\left(-\alpha|x|^{2}-\beta|\xi|^{2}\right)
\end{displaymath}
then
\begin{displaymath}
l^{\#}_{1}(t)=f_{0}\exp\left(-\int^{t}_{0}R^{\#}(u_{0})(\tau)d\tau\right)\;\;\;\;\mathrm{and}\;\;\;\;u^{\#}_{1}(t)=f_{0}+\int^{t}_{0}Q^{\#}_{+}(u_{0},u_{0})(\tau)d\tau,
\end{displaymath}
clearly $0\leq l^{\#}_{0}\leq l^{\#}_{1}\leq u^{\#}_{1}$.  In addition, from the last expression and estimate (\ref{E3ECO})
\begin{displaymath}
u^{\#}_{1}(t)\leq \left(\left\|f_{0}\right\|_{\alpha,\beta}+k_{\alpha,\beta}\left\|u^{\#}_{0}\right\|^{2}_{\alpha,\beta}\right)\exp\left(-\alpha|x|^{2}-\beta|\xi|^{2}\right).
\end{displaymath}
Note that $\left\|u^{\#}_{0}\right\|_{\alpha,\beta}=C$, therefore it suffices to choose $C$ such that
\begin{displaymath}
\left\|f_{0}\right\|_{\alpha,\beta}+k_{\alpha,\beta}C^{2}=C
\end{displaymath}
to satisfy the beginning condition.  This is possible as long as
\begin{displaymath}
\left\|f_{0}\right\|_{\alpha,\beta}\leq\frac{1}{4k_{\alpha,\beta}},
\end{displaymath}
this concludes the proof of the theorem.
\end{proof}
\begin{Theorem}\label{ET2}
Under the same conditions of Theorem~\ref{ET1} the problem (\ref{E8LP}) has a unique weak solution in the space
\[
\mathcal{A}=\left\{f:f^{\#}\in L^{\infty}(0,T;M^{\alpha,\beta}_{+})\right\}.
\]
\end{Theorem}
\begin{proof}
Note that the mild solution $f^{\#}$ of problem (\ref{E8LP}) lies in $L^{\infty}(0,T;M^{\alpha,\beta}_{+})$.  As a result, $Q(f,f)$ lies in $L^{1}(0,T;L^{1})$.  Therefore, one can use Theorem~\ref{AT1} to prove that $f$ is in fact a weak solution of (\ref{E8LP}).\\
The uniqueness follows because if $f$ and $g$ are weak solutions that lie in $\mathcal{A}$ then $Q(f,f)$ and $Q(g,g)$ lie in $L^{1}(0,T;L^{1})$ hence by Theorem~\ref{AT1} they are mild solutions, thus invoking Theorem~\ref{ET1} it is concluded that $f=g$.
\end{proof}
\noindent $Remarks:$ First, it is important to observe that the conservation of mass and loss of energy properties ($P1_{a}$) and ($P1_{b}$) were never used in order to prove Theorem~\ref{ET1}.  In this sense there is no hope to prove large data existence of solutions with this technique unless these properties are included in the argument.  Indeed, in the theory of \textsl{Renormalized solutions} for elastic BE properties ($P1_{a}$) and ($P1_{b}$) are essential (in addition to the bounded entropy property).\\\\
Second, Theorem~\ref{ET1} shows that there is no clustering (accumulation of mass) in the solution for small data as long as the restitution coefficient satisfies the conditions \textsl{(A1), (A2) and (A3)} in section (2).  Observe also that since $f^{\#}\in L^{\infty}(0,\infty;M^{\alpha,\beta}_{+})$ one has
\begin{displaymath}
0\leq f\leq C\exp\left(-\alpha|x-t\xi|^{2}\right)\exp\left(-\beta|\xi|^{2}\right)
\end{displaymath}
for $C>0$ given in Theorem~\ref{ET1}.  Therefore,
\begin{displaymath}
\lim_{t\rightarrow\infty}f(t,x,\xi)=0\;\;\;\mathrm{a.e.\;\;in\;\;\mathbb{R}^{n}\times\mathbb{R}^{n}}.
\end{displaymath}
Thus, for sufficiently small data the equilibrium is always the trivial one.  Similarly as Illner $\&$ Shinbrot pointed out in \cite{Il}, it is possible to estimate the asymptotic decay for the spatial density
\begin{displaymath}
\rho(t,x)=\int_{\mathbb{R}^{n}}f(t,x,\xi)d\xi.
\end{displaymath}
Indeed, note that
\begin{displaymath}
\rho(t,x)\leq C\int_{\mathbb{R}^{n}}\exp\left(-\alpha|x-t\xi|^2-\beta|\xi|^{2}\right)d\xi\leq \frac{C}{t^{n}}\int_{\mathbb{R}^{n}}\exp\left(-\alpha|z|^2\right)dz 
\end{displaymath}
whence,
\begin{displaymath}
\rho(t,x)=o\left(\frac{1}{t^{n}}\right)\;\;\mathrm{as}\;\;\;t\rightarrow+\infty.
\end{displaymath}
Finally, observe that thanks to the properties $(P1_{a})$ and $(P1_{b})$ of the collision operator, the mild-weak solution $f$ has the following usual properties for $t\geq 0$:
\begin{description}
\item [\it (1)]Conservation of mass and momentum
\[
\int_{\mathbb{R}^{n}}\int_{\mathbb{R}^{n}}f(t)(1+\xi)d\xi dx=\int_{\mathbb{R}^{n}}\int_{\mathbb{R}^{n}}f_{0}(1+\xi)d\xi dx\;\;\mbox{and,}
\]
\item [\it (2)]Dissipation of energy
\[
\int_{\mathbb{R}^{n}}\int_{\mathbb{R}^{n}}f(t)\;|\xi|^{2}d\xi dx\leq\int_{\mathbb{R}^{n}}\int_{\mathbb{R}^{n}}f_{0}\;|\xi|^{2}d\xi dx.
\]
\end{description}
In fact, the solution $f$ has all moments bounded due to the Maxwellian decay in velocity.

%%%%%%%%%%%%%%%%%%%%%%%%%%%%%%%%%%%%%%%%%%%%%%%%%%%%%%%%%%%%%%%%%%%%%%%%%%%%%
%%%%%%%%%%%%%%%%%%%%%%%%%%%%%%%%%%% Appendix %%%%%%%%%%%%%%%%%%%%%%%%%%%%%%%%
%%%%%%%%%%%%%%%%%%%%%%%%%%%%%%%%%%%%%%%%%%%%%%%%%%%%%%%%%%%%%%%%%%%%%%%%%%%%%

\appendix

\section{Some facts for the mild and weak solutions for the Inhomogeneous Boltzmann Problem}
This section includes some elementary results that are needed along the paper.  Let us start with a well known result about the spaces $W^{1,p}(0,T;X)$.  Its proof can be found in \cite{Evans}.
\begin{Theorem}\label{T2.5}
Let $X$ a Banach space, and let $f\in W^{1,p}(0,T;X)$ for some $1\leq p\leq+\infty$.  Then
\begin{description}
\item [\it (1)]$f\in C(0,T;X)$ (after possibly being redefined on a set of measure zero), and
\item [\it (2)]
\[
f(t)=f(s)+\int^{t}_{s}f_{t}(\tau)d\tau\;\;\mbox{for all}\;\;0\leq s\leq t\leq T.
\]
\end{description}
\end{Theorem}
\noindent In the same way that the mild solution was defined, one can define a different concept of solution for a transport problem, namely, the weak solution.  Let $T:=\partial_{t}+\xi\cdot\nabla$ the transport operator, $\Omega\subset\times\mathbb{R}^{n}\times\mathbb{R}^{n}$ an open set, and $h\in L^{1}(0,T;L^{1}_{loc}(\Omega))$.  Then one can look for a function $f$ satisfying the transport equation
\begin{align}
Tf&=h\;\mbox{in}\;(0,T)\times\Omega\nonumber\\
f&=f_{0}\;\mbox{on}\;\{0\}\times\Omega\label{Dweak}
\end{align}
in the following sense,
\begin{Definition}\label{defweak}
A function $f\in L^{1}(0,T;L^{1}_{loc}(\Omega))$ is called weak solution of problem (\ref{Dweak}) if for any $\psi\in D((0,T)\times\Omega)$
\begin{description}
\item [\it (1)]
\[
-\int^{T}_{0}\int_{\Omega}f\;T\psi=\int^{T}_{0}\int_{\Omega}h\;\psi\;\;\mbox{and,}
\]
\item [\it (2)]$f(0,\cdot)=f_{0}\;\;\mbox{a.e. in}\;\;\Omega$.
\end{description}
\end{Definition}
\noindent The following is a classical result in linear transport that relates mild solutions and weak solutions
\begin{Theorem}\label{AT1}
Take $\Omega=\mathbb{R}^{n}\times\mathbb{R}^{n}$ in problem (\ref{Dweak}), and assume that $f_{0}\in L^{1}$ and $h\in L^{1}(0,T;L^{1})$.  Then, $f$ is a weak solution if and only if $f$ is a mild solution for this problem.
\end{Theorem}
\begin{proof}
Assume that $f$ is a weak solution for problem (\ref{Dweak}).  Thus, for any $\psi\in D((0,T)\times\Omega)$ 
\[
-\int^{T}_{0}\int_{\Omega}f\;T\psi=\int^{T}_{0}\int_{\Omega}h\;\psi.
\]
Let $\sigma\in D(\Omega)$ and $\rho\in D((0,T))$.  Take $\psi(t,x,\xi)=\rho(t)\sigma(x-t\xi,\xi)$ in this equation and perform the change the change of variables $(t,x,\xi)\rightarrow (t,x+t\xi,\xi)$ to obtain
\begin{displaymath}
-\int^{T}_{0}\int_{\Omega}f^{\#}(t)\rho'(t)\psi(x,\xi)=\int^{T}_{0}\int_{\Omega}h^{\#}(t)\rho(t)\psi(x,\xi).
\end{displaymath}
This works for any $\psi$, hence
\[
-\int^{T}_{0}f^{\#}\rho'(t)=\int^{T}_{0}h^{\#}(t)\rho(t)dt\;\;\;\mbox{a.e. in}\;\;\Omega.
\]
Therefore, using the definition of weak derivative one concludes that
\begin{equation}\label{A4}
\frac{df^{\#}}{dt}=h^{\#}\in L^{1}(0,T;L^{1}).
\end{equation}
Recall that $f\in L^{1}(0,T;L^{1}_{loc})$, whence same property holds for $f^{\#}$.  So for any compact set $K\subset\Omega$, one applies Theorem~\ref{T2.5} to obtain that $f^{\#}\in C(0,T;L^{1}_{loc})$, and that for all $t\geq0$ the equation
\begin{equation}\label{A5}
f^{\#}(t)=f_{0}+\int^{t}_{0}h^{\#}(\tau)d\tau\;\;\mbox{in}\;\;L^{1}_{loc}
\end{equation}
holds.  Since the right hand side of equation (\ref{A5}) belongs to $L^{1}(0,T;L^{1})$, it must be that at each time $t$ equation (\ref{A5}) holds in fact in $L^{1}$, therefore, $f^{\#}\in L^{1}(0,T;L^{1})$.  As a result, $f\in W^{1,1}(0,T;L^{1})$.\\\\
For the converse use equation (\ref{A4}) and proceed backwards using the same idea.
\end{proof}

\Ack{I would like to thank Irene Gamba and Vlad Panferov for their guidance and fruitful discussions.  The partial support from the NSF under grant DMS-0507038 is also gratefully acknowledged.}

%%%%%%%%%%%%%%%%%%%%%%%%%%%%%%%%%%%%%%%%%%%%%%%%%%%%%%%%%%%%%%%%%%%%%%%%%%%%%
%%%%%%%%%%%%%%%%%%%%%%%%%%%%%%%%%%% Bibliography %%%%%%%%%%%%%%%%%%%%%%%%%%%%
%%%%%%%%%%%%%%%%%%%%%%%%%%%%%%%%%%%%%%%%%%%%%%%%%%%%%%%%%%%%%%%%%%%%%%%%%%%%%

%%%%%%%%%%%%%%%%%%%%%%%%%%%%%%%%%%%Signature%%%%%%%%%%%%%%%%%%%%%%%%%%%%%%%%%

\signra         

\end{document}